\documentclass[12pt]{article}

\usepackage{amsmath, euscript, amssymb, amsfonts}

\newcommand{\oR}{{\mathbb R}}
\newcommand{\oV}{{\mathbb V}}
\newcommand{\oC}{{\mathbb C}}
\newcommand{\oT}{{\mathbb T}}

\newcommand{\oW}{{\mathbb W}}

\begin{document}
\sloppy

\title{NONLOCAL EXTENSION OF THE BORCHERS CLASSES OF QUANTUM FIELDS}

\author{M. A. Soloviev}
\date{}
\maketitle

\begin{center}
{\small I. E. Tamm Department of Theoretical Physics, P. N.
Lebedev Physical Institute, Leninsky prosp. 53, Moscow 119991,
Russia, e-mail: soloviev@lpi.ru}
\end{center}

\bigskip

\hfill         {\it To the fond memory of Professor M.~S.~Marinov}
\vspace{1cm}
\begin{center}
{\bf Abstract}
\end{center}

We formulate an equivalence relation between
nonlocal quantum fields,
generalizing the relative locality which was studied by Borchers in the
framework of local QFT. The Borchers classes are shown to allow a natural
extension involving nonlocal fields with arbitrarily singular ultraviolet
behavior. Our consideration is based on the systematic employment of the
asymptotic commutativity condition which, as established previously, ensures
the normal spin and statistics connection as well as the existence of {\it
PCT} symmetry in nonlocal field theory.  We prove the transitivity of the
weak relative asymptotic commutativity property generalizing Jost-Dyson's
weak relative locality and show that all fields in the same extended Borchers
class have the same S--matrix.

 \section{Introduction}

In this article, some new results about the possibility of generalizing the
 {\it PCT} theorem to nonlocal interactions are reviewed and applied
 to the corresponding extension of the Borchers classes of quantum
 fields.  Nonlocal QFT's were extensively studied in the 1970s with the hope
 of overcoming the problems of non-renormalizable Lagrangian field theories.
 At that time, I was fortunate to work in ITEP together with M.~S.~Marinov
 and I never forget the friendly encouragement of this outstanding person.
 Gauge theories and constrained dynamics were the main subjects of
 discussions at a seminar organized by M.~S.~Marinov for our
 mathematical group. At the same time, I tried extending the scattering
 theory of particles to nonlocal interactions.  This direction of research
 was initiated by another eminent scientists Professor N.~N.~Meiman~\cite{M},
 who also passed away recently.

 At present nonlocal QFT models are interesting first of all
 in connection with string theory and D--brane theory. This connection is
 best seen from the holographic point of view~\cite{G}.  Bounds for the
 S-matrix derivable from nonlocal field theories could provide a way of
 investigating the bulk locality properties in the AdS/CFT correspondence
 after extracting flat-space scattering amplitudes from the boundary
 conformal field theory correlators.  On the other hand, the question of a
 possible {\it PCT} invariance violation caused by nonlocality is crucial for
 phenomenological schemes exploiting propagators with nonlocal form-factors
 suppressing ultraviolet divergences and proposed as an alternative to string
 theory, see e.g.,~\cite{M1}.  It is equally
 important to characterize the nonlocal theories that have the same S-matrix
 as usual ones. The present status of this problem first raised in~\cite{TC}
 and related topics is just what we will discuss below.

The article is organized as follows. Section 2 contains a brief sketch of
those tools of modern functional analysis that make it possible to
extend the basic results of axiomatic approach~\cite{SW,BLOT} to nonlocal
interactions.  Next we formulate an asymptotic commutativity condition
which replaces local commutativity and is nearer macrocausality.  In Section
3, we outline the proof of a new uniqueness theorem~\cite{S1,S2} for
distributions which plays a central role in deriving the  {\it PCT} theorem
and the spin-statistics relation for nonlocal quantum fields. The significance
of the notion of analytic wave front set to these derivations is explained.  In
Section 4, we describe some important properties of vacuum expectation values
of nonlocal fields which follow from the Lorentz covariance, the spectral
condition, and from the fact that the complex Lorentz group contains the
total space-time inversion.  In Section 5, a condition of weak asymptotic
commutativity is defined and its equivalence to the existence of {\it PCT}
symmetry is demonstrated.  In Section 6, we prove the transitivity of the
weak relative asymptotic commutativity and show that this property  leads to
a natural extension of the Borchers classes of quantum fields.  In Section 7,
we argue that all local and nonlocal fields belonging to the same extended
Borchers class have the same S--matrix.  Section 8 contains concluding
remarks.

 \section{Carrier cones of analytic functionals and \\ asymptotic
 commutativity of nonlocal fields}

  We will consider field theories in which correlation functions can be so
  singular in their space-time dependence that these singularities violate
 locality. It is well known~\cite{M} that a breakdown of local
 commutativity occurs if the Fourier transforms of correlation functions have
 an exponential growth of order $\geq 1$ and, among a variety of nonlocal
 theories, these seem to be most closely related to string theory which is
 characterized by an exponentially increasing density of states,
 see~\cite{K,AB} for more detailed comments. From a technical point of view,
 this means that we abandon the usual assumption~\cite{SW,BLOT} that the
 vacuum expectation values of products of fields are tempered distributions
 defined on the Schwartz space  $S$ consisting of infinitely differentiable
 functions of fast decrease. The expectation values are supposed instead to
 be well defined on test functions analytic in coordinate space, i.e., are
 regarded as analytic functionals. Then observables emerges only above a
 definite length scale, but an analysis~\cite{S3,S4} shows that a large class
 of analytic functionals retain a kind of angular localizability.
 This is just the property that enables one to develop a self-consistent
 scattering theory for nonlocal interactions.
 In~\cite{S3,S4}, we use the spaces  $S^0_\alpha$
 introduced by Gelfand and Shilov~\cite{GS}, which are most suitable for
 nonperturbative formulation of indefinite metric QFT's and particularly,
 gauge theories.  In nonlocal field theories satisfying the
 positivity condition, another space  $S^0=S^0_\infty$ is commonly employed.
 The latter is none other than the Fourier transform of the space
  ${\cal D}$ consisting of infinitely differentiable functions of compact
  support, and this choice of test functions implies that the expectation
  values have a finite order of singularity in momentum space.
   The methods and results
  of~\cite{S3,S4} can be extended to cover analytic functionals
  defined on $S^0$ but this is not a trivial exercise because of some
  topological complications.  Nevertheless, we outline this extension in view
  of the important role of the space ${\cal D}$ in the theory of
  distributions.

To each open cone $U\subset \oR^n$, we assign a space
$S^0(U)$ consisting of those entire analytic functions on $\oC^n$,
that satisfy the inequalities
$$
|f(z)| \le C_N\left(1+\|x\|\right)^{-N}e^{b\|y\|+bd(x,U)} \quad (N=0,1,\dots),
 \eqno{(1)}
$$
where $b$ and $C_N$ are positive constants depending on $f$, and
$d(\cdot,U)$ is the distance from the point to the cone
$U$.  (The norm in $\oR^n$ is assumed to be Euclidean in what follows.)
This space can naturally be given a topology by regarding it as the inductive
limit of the family of countably normed spaces $S^{0,b}(U)$ whose norms are
defined in accordance with the inequalities (1), i.e.,
 $$
 \|f\|_{U,b,N} =
\sup_z |f(z)|\left(1+\|x\|\right)^N e^{-b\|y\|-bd(x,U)}.
 \eqno{(2)}
 $$
For each closed cone $K\subset \oR^n$, we also define a space $S^0(K)$
by taking another inductive limit through those open cones $U$ that contain
the set $K\setminus\{0\}$ and shrink to it. Clearly,
 $S^0(\oR^n)=S^0$. As usual, we use a prime to denote the continuous dual of
 a space under consideration.
   A closed cone $K\subset\oR^n$
is said to be a {\it carrier} of a  functional $v\in S^{\prime 0}$
  if $v$ has a continuous extension to the
space $S^0(K)$, i.e., belongs to $S^{\prime 0}(K)$.
  We refer to~\cite{S3,S4}
for a motivation of this definition and for its connection with  the
Sato-Martineau theory of hyperfunctions. As is seen from estimate (1),
 this property may be thought of as a fast decrease (no worse
than an exponential decrease of order 1 and maximum type)  of $v$ in the
complement of $K$. It should also be emphasized that  if  $v$ is a tempered
distribution with support in $K$, then the restriction $v|S^0$ is carried by
 $K$.

 We list the basic facts which allow handling the analytic
 functionals of class  $S^{\prime 0}$, in most cases, as easily as tempered
distributions.

 1.  {\it The spaces $S^0(U)$ are Hausdorff and complete.
 A  set $B\subset S^0(U)$ is bounded if and only if it is contained
in some space $S^{0,b}(U)$ and is bounded in each of its norms.}

The proof given in ~\cite{S5}  relies on the acyclicity  of the
 injective sequence of Fr\'echet spaces $S^{0,b}(U)$.

 2. {\it The space $S^0$ is dense in every $S^0(U)$ and in every $S^0(K)$}.

As shown in~\cite{S1}, this follows from an analogous theorem proved for
 $S^0_\alpha$ in~\cite{S4}.

 3. {\it If a functional $v\in S^{\prime 0}$ is carried by each of closed
  cones $K_1$ and $K_2$, then it is carried by their intersection}.

 Because of this, there is a smallest $K$ such
 that $v\in S^{\prime 0}(K)$. The proof is similar to that given
 for $S^{\prime 0}_\alpha$ in~\cite{S3}, but appeals besides
 to the topological Lemma  5.11 in~\cite{S6}.

 4. {\it If $v\in S^{\prime 0}(K_1\cap K_2)$, then $v=v_1+v_2$, where
$v_j\in S^{\prime 0}(U_j)$ and $U_j$ are any open cones such that}
$U_j\supset K_j\setminus\{0\}$, $j=1,2$.

This theorem can also be proved in a manner similar to that of~\cite{S3},
 but  using auxiliary conic neighborhoods with additional regularity
properties specified in~\cite{S6}.  Within the framework of functional spaces
 $S^{\prime 0}_\alpha$, a stronger decomposition property holds with
  $v_j$ carried by the cones $K_j$ themselves; such an improvement is also
possible for $S^{\prime 0}$ if $K_1\cap K_2=\{0\}$, see  Theorem 5
in~\cite{S5}. For applications to QFT, the case of a properly convex cone
 $K$ is of special interest. Then the dual cone $K^*=\{\eta:\,\eta
x\geq0,\quad \forall x\in K\}$ has a nonempty interior and $e^{i\zeta
x}\in S^0(K)$ for all ${\rm Im}\,\zeta\in {\rm int}\,K^*$. Therefore, the
Laplace transformation
  $$
  \check {\bf v}(\zeta)=(2\pi)^{-n}(v,\,e^{i\zeta x})
   \eqno{(3)}
  $$
  is well defined for each $v\in S^{\prime 0}(K)$. It is easily verified that
  the function (3) is analytic in the tubular domain
 $\oR^n+i\,{\rm int}\,K^*$ and satisfies the estimate
$$
 |\check {\bf v}(\zeta)|\,\leq\,C_ {R,V^\prime}\,
|{\rm Im}\,\zeta |^{-N_R}\qquad ({\rm Im}\,\zeta\in V^\prime,\ |\zeta|\leq R)
\eqno{(4)}
$$
for any $R>0$ ¨ for each cone $V^\prime$ such that
 $\overline{V}^\prime\setminus\{0\}\subset {\rm int}\,K^*$. What is more, the
above results make it possible to establish the following theorem of the
Paley-Wiener-Schwartz type.

 5. {\it For every properly convex closed cone $K\subset \oR^n$,
 the Laplace transformation isomorphically maps  $S^{\prime
 0}(K)$ onto the space of functions analytic in the tube
  $\oR^n+i\,{\rm int}\,K^*$ and satisfying $(4)$. The Fourier transform of
  $v$ is a boundary value of the function $(3)$, i.e., $\check {\bf
 v}(\zeta)$ converges to $\check v$ in  ${\cal D}'$ as ${\rm Im}\,\zeta\to 0$,
   ${\rm Im}\,\zeta\in V^\prime $.}

We will consider a finite family of fields
$\{\phi_\iota\}$ that are operator-valued generalized functions defined
   on the test function space $S^0(\oR^4)$
and transform  according to irreducible representations of the proper
Lorentz group $L_+^\uparrow$ or its covering group $SL(2,\oC)$.  We
   adopt all the standard assumptions of the Wightman axiomatic
   approach~\cite{SW,BLOT} except local commutativity  which
   cannot be formulated in terms of the analytic test functions.
   It should be noted that, using $S^0$, we do not impose any
   restrictions on the high-energy (ultraviolet) behavior of fields because
   the test functions are of compact support in momentum space.
   In this sense, the space $S^0$ is universal for nonlocal fields. We
   denote by $D_0$ the minimal common invariant domain, which is assumed to
   be dense, of the field operators in the Hilbert space ${\cal H}$ of states,
   i.e., the vector subspace of ${\cal H}$ that is spanned by the vacuum state
   $\Psi_0$ and by various vectors of the form
    $$
   \phi_{\iota_1
   \ell_1}(f_1)\dots \phi_{\iota_n \ell_n}(f_n)\Psi_0 \qquad(n=1,2,\dots),
    $$
   where  $f_k \in S^0(\oR^4)$ and $\ell_k$ are the Lorentzian indices.
   The space $S^0$, being Fourier-isomorphic to ${\cal D}$, is nuclear.
   Therefore, the  $n$-point vacuum expectation values  uniquely
   determine  Wightman generalized functions ${\cal
  W}_{\iota_1 \ell_1,\dots,\iota_n \ell_n}\in S^{\prime 0}(\oR^{4n})$
  and we identify these objects, just as in the standard
  scheme~\cite{SW,BLOT}.  The property of nuclearity allows us to define as
   well the expressions
       $$
       \int \phi_{\iota_1 \ell_1}(x_1)\dots \phi_{\iota_n
     \ell_n}(x_n)\, f(x_1,\dots,x_n)\,{\rm d}x_1\dots{\rm d}x_n \Psi_0
     \qquad(n=1,2,\dots),
       \eqno{(5)}
       $$
      where $f \in S^0_\alpha(\oR^{4n})$, and to verify that
         every operator $\phi_{\iota\ell}(f)$ can be extended to the subspace
       $D_1\supset D_0$ spanned by vectors (5).

      The foregoing inspires  the following definition.

     {\bf Definition 1}. The field components
  $\phi_{\iota\, \ell}$ and  $\phi_{\iota^\prime \ell^\prime}$
 commute (anticommute) asymptotically for large spacelike separation of
 their arguments if the functional
   $$ \langle\Phi,\, [\phi_{\iota\,\ell}(x),\phi_{\iota^\prime
\ell^\prime}(x')]_{\stackrel{-}{(+)}}\Psi\rangle
  \eqno{(6)}
  $$
is carried by the cone $\overline{\oW}=\{(x,x') \in \oR^8:  (x-x')^2\geq 0\}$
for any vectors $\Phi, \Psi\in D_0$.

   We replace the local commutativity axiom  by the
   {\it asymptotic commutativity} condition which means that any two field
components either commute or anticommute asymptotically.
This condition is evidently weaker than local commutativity in the sense that
it is certainly  fulfilled for the restrictions of local tempered
distribution fields to $S^0(\oR^4)$. The standard considerations of Lorentz
covariance imply that the type of the commutation relation depends only on
the type of the participating fields, not on their Lorentzian indices, and we
drop these indices in what follows.

   Lemma 3 in~\cite{S2} shows that under the stated condition, the functional
  $$ {\cal
 W}_{\iota_1,\ldots,\iota_n}(x_1,\ldots,x_k,x_{k+1},\ldots,x_n)- {\cal
W}_{\,\iota_1,\ldots,\iota_{k+1},\iota_k,\ldots,\iota_n}
(x_1,\ldots,x_{k+1},x_k,\ldots,x_n),
  \eqno{(7)}
   $$
    where the sign $-$ or $+$ corresponds to the type of commutation relation
  between $\phi_{\iota_k}$ and  $\phi_{\iota_{k+1}}$, is carried by the cone
$\overline{\oW}_{n,k}=\{ x\in \oR^{4n}:\,(x_k-x_{k+1})^2\geq 0\}$.
   It follows, in particular, that if the asymptotic commutativity condition
   is satisfied for $\Phi, \Psi\in D_0$, then it is also satisfied for
   $\Phi, \Psi\in D_1$. Moreover, it is fulfilled for any
   $\Phi\in {\cal H}$, $\Psi\in D_1$. In other words, $\overline{\oW}$ is a
   carrier of the vector-valued functional $$ \Xi(f)=\int[\phi_{\iota}(x),
   \phi_{\iota^\prime}(x')]_{\stackrel{-}{(+)}} f(x,x')\,{\rm d}x {\rm
   d}x'\,\Psi\qquad (\Psi\in D_1)
  \eqno{(8)}
  $$
  defined on $S^0(\oR^8)$. Indeed, $\|\Xi(f)\|^2={\cal B}(\bar f,f)$, where
  ${\cal B}$ is a separately continuous bilinear form on $S^0(\oR^8)$ with
  the property that both linear functionals determined by it when one of its
  two arguments is held fixed are carried by $\overline{\oW}$.
   A consideration analogous to that of
  Lemma 3 in~\cite{S2} shows that any bilinear form possessing this property
  is identified with an element of the space $S^{\prime 0}(\overline{\oW}
  \times \overline{\oW})$.  The main point of the argument is the equality
  $S^0(U\times \oR^d)=S^0(U) \mathbin{\hat{\otimes}_i}S^0(\oR^d)$, where the
  tensor product is equipped with the inductive topology and the hat means
  completion.  If $f_\nu\in S^0(\oR^8)$ and $f_\nu\to f\in
  S^0(\overline{\oW})$, then $(\bar f_\nu -\bar f_\mu)\otimes (f_\nu -
   f_\mu)\to 0$ in the topology of $S^0(\overline{\oW} \times
   \overline{\oW})$ as $\nu, \mu\to \infty$.  Therefore, the sequence
   $\Xi(f_\nu)$ converges strongly in ${\cal H}$ and functional (8) has a
   continuous extension to $S^0(\overline{\oW})$.

\section{Analytic wave front set and carriers}

The classical derivation of the spin-statistics relation and {\it PCT}
symmetry in local QFT~\cite{SW,BLOT} is based on exploiting the analytic
properties of vacuum expectation values in $x$-space and substantially employs
a uniqueness theorem of complex analysis which asserts that if a function
${\bf w}(z)$ $(z=x+iy\in \oC^n)$ is
analytic on a tubular domain whose basis is an open connected cone and if its
boundary value on $\oR^n$ vanishes in a nonempty open set, then ${\bf w}$ is
identically zero. By using this theorem, it is easy to derive another
 uniqueness theorem which shows that if a tempered distribution $u\in S'$
has support in a properly convex cone $V$ and its Fourier transform
 $\hat u$ vanishes on an open set, then $u\equiv 0$.
Indeed, by the Paley-Wiener-Schwartz theorem (see, e.g., Theorem
2.9 in~\cite{SW}), the support condition implies the existence of the Laplace
transform $\hat {\bf u}(z)=(u,\,e^{-(\cdot,z)})$ analytic in the tube
 $\oR^n-i\,{\rm int}\, V^*$ and  whose boundary value is $\hat u$.
 In~\cite{S1,S2}, a natural analog of the latter uniqueness property is
 established for distributions in ${\cal D}'$, which provides a means for the
 extension of the spin-statistics and  {\it PCT} theorems to nonlocal fields
 with arbitrary high-energy behavior.

  {\bf Theorem 1}. {\it Let $u\in {\cal D}^{\prime}(\oR^n)$ be a
  distribution whose   support is contained in a properly convex cone $V$.
  If its Fourier transform   $\hat u\in S^{\prime 0}(\oR^n)$
  is carried by a cone different from $\oR^n$, then $u\equiv 0$.}

  We will not repeat the proof given in~\cite{S1,S2}, but we
  explain the role that is played by the notion of analytic wave front set
   $WF_A(u)$ in the derivation of this theorem.  Recall~\cite{H}, that
   $WF_A(u)$ consists of pairs $(p,\xi)$, where $p$ ranges the smallest
   closed subset of $\oR^n$ outside of which the distribution $u$ is
   analytic. This subset is denoted by ${\rm  sing\,supp}_A u$, and for every
   $p\in {\rm sing\,supp}_A u$, the vector $\xi$ ranges the closed cone in
    $\oR^n\setminus \{0\}$ formed by those directions of a ``bad'' behavior
    of the Fourier transform $\hat u$ at infinity that are responsible for
     the nonanalyticity of $u$ at the point $p$.  If a closed cone $K\subset
     \oR^n$ is a carrier of $\hat u$, then~\footnote{It should be noted that
 the operator $u\to \hat u$ is dual of the test function transformation
 $f(x)\to \int  e^{-ipx}f(x)\,{\rm}dx$, with the minus sign in the exponent,
  as opposite to $v\to \check v$.} $$ WF_A(u)\subset \oR^n \times(K\setminus
  \{0\}).  \eqno{(9)} $$ In other words, the directions external to carrier
  cone cannot be responsible for singularities of $u$.  Formula (9) improves
  considerably Lemma 8.4.17 in~\cite{H} which asserts that, for each tempered
   distribution $u\in S^\prime$, the inclusion  $WF_A(u)\subset \oR^n \times
  (L\setminus \{0\})$ holds, where $L$ is the limit cone of the set ${\rm
  supp}\,\hat u$ at infinity.  This cone consists of the limits of various
  sequences $t_\nu x_\nu$, where $x_\nu\in {\rm   supp}\,\hat u$ and
  $0<t_\nu\rightarrow 0$, and it is certainly a  carrier of the restriction
  $\hat u|S^0$.  The derivation of (9) is based on the employment of the
  decomposition theorem and the Paley-Wiener-Schwartz-type theorem stated in
  Section 2. By using this inclusion, we can easily prove Theorem 1 in the
  simplest case when $0\in {\rm   supp}\,u$.  Then every vector in
  the cone $-V^*\setminus\{0\}$ is an external normal to the support at the
  point 0.  By Theorem 9.6.6 of~\cite{H},  all the nonzero elements of the
  linear span of external normals belong to $WF_A(u)_{p=0}$.
  Because the cone $V$ is properly convex, the interior of $V^*$ is not
  empty, and this linear span covers $\oR^n$. Therefore, each carrier cone of
  $\hat u$ must coincide with $\oR^n$. The general case can be reduced to
  this special case by considering the series  of "contracted"
   ultadistributions
   $$
   \sum^\infty_{\nu=1}c_\nu u_\nu,\qquad (u_\nu,g(p))
  \stackrel{\rm def}{=}\nu^{-n}(u,g(p/\nu)).
   \eqno{(10)}
   $$
  As shown in~\cite{S1,S2}, the coefficients $c_\nu$ can be chosen such
  that this series converges in ${\cal D}^{\prime}$ to a distribution whose
  support contains the point 0 and whose Fourier transform is carried by the
  same cones that  $\hat u$ is.

  We present here another result related to Theorem 1 and concerning Fourier
  hyperfunctions.  Recall that this name is used for the continuous linear
   functionals defined on the Fourier-symmetric space $S^1_1$
   of functions analytic in an complex $l$--neighborhood
  of the real space and satisfying the estimate $|f(z)|\leq C\,e^{b\|z\|}$,
  where $l$, $b$, and $C$ depend on  $f$.  Such functionals, as well as their
  Fourier transforms, have uniquely defined supports and are used in the
  most general formulation of local QFT, see~\cite{BN1}.

    {\bf Theorem 2}. {\it Let $u\in S^{\prime 1}_1(\oR^n\times \oR^m)$ and
    ${\rm supp}\,u \subset V\times \oR^m$, where $V$ is a properly convex
    cone in $\oR^n$.  Then ${\rm supp}\,\hat u= \oR^n\times
    M$, where $M$ is a closed set in $\oR^m$.}

    An analogous theorem holds evidently for tempered distributions.
    Indeed, if $x\not\in {\rm supp}\,\hat u$, then there is a neighborhood of
     $x$ of the form $U_1\times U_2$ and not meeting ${\rm supp}\,\hat u$.
    Let $u\in S'$ and $f\in {\cal D}(U_2)$. Then the distribution $\int
    \hat u(x_1,x_2)\, f(x_2)\,{\rm d}x_2$ vanishes in $U_1$ and its
    (inverse) Fourier transform has support in the convex cone $V$.
    Therefore, this distribution  is identically zero and we conclude that
    ${\rm supp}\,\hat u$ does not meet $\oR^n\times U_2$ because any test
    function localized in this region can be approximated by functions
    belonging to ${\cal D}(\oR^n)\otimes{\cal D}(U_2)$.

    This argument is unapplicable to Fourier hyperfunctions because
     $S^1_1$ does not contain any functions of compact support.
     However, we may make use of the inclusions
   $$
     N({\rm supp}\,\hat u)\subset WF_A(\hat u)\subset \oR^{n+m}\times
     (V\setminus \{0\}),
   \eqno{(11)}
  $$
  where $N(\cdot)$ is the set of normals.
     The first of inclusions (11) holds according to~\cite{H}, and the second
  can be proved in complete analogy to the derivation of (9), because a
   suitable decomposition theorem and a Paley-Wiener-Schwartz-type theorem
   are well known for  Fourier hyperfunctions, see Propositions 2.7 and 2.9
   in~\cite{BN1}.  The set $N(\cdot)$ includes both external and internal
   normals, whereas the cone $V$ is properly convex and does not contain a
   straight line.  Therefore, (11) implies that  any normal to the
   closed set ${\rm supp}\,\hat u$ has zero projection on $\oR^n$ and so
   this set must be of the form indicated above.

   A few words of explanation concerning the last conclusion are, perhaps,
 necessary. Let $X$ be a closed set in $\oR^n\times \oR^m$, $y\not\in X$, and
  let $z$ be a point in $X$ with minimal distance to $y$. Then $y-z\in
  N(X)|_z$ by the definition~\cite{H}. We know that for any
   $y$, the projection of $y-z$ on $R^n$ is zero and we need to show that
  then $x=(x_1,x_2)\in X$ implies $x'=(x'_1,x_2)\in X$ for any
  $x'_1\in \oR^n$.  Assume contrarily that there is a point
   $x'\not\in X $ of this form. For convenience, we
   suppose that $x_2=0$ and identify $\oR^n$ with its canonical image in
  $\oR^n\times\oR^m$. Let $B$ be the open ball in $\oR^n$ which is centred at
   $x'_1$ and  whose radius is equal to the distance from $x'_1$ to
   $X\cap\oR^n$.  For every $y\in B$, let $r(y)>0$ be the squared distance of
   $y$ from $X$. By our assumption, any point $z\in X$ such that
  $\|y-z\|^2=r(y)$ is of the form $z=(y, z_2)$ and hence
  $\|z_2\|^2=r(y)$.  Let $z'\in X$ be such that  $\|y'-z'\|^2=r(y')$.
  Then $r(y)\leq \|y-z'\|^2=\|y-y'\|^2+r(y')$.
   Analogously, $r(y')\leq \|y-y'\|^2+ r(y)$.  Therefore,
  the function $r(y)$ is differentiable and its derivative is identically
  zero, i.e., this function is constant, but this contradicts that it must
  tends to zero as $y$ nears a boundary point of $B$ belonging to $X$.
  This completes the proof.

  \section{Jost points in  nonlocal QFT}

 In local field theory, Jost points are
 real points of the extended domain of analyticity of the Wightman
functions  ${\cal W}(x_1,\dots,x_n)$. The Bargman-Hall-Wightman theorem shows
that this extension is obtained by applying various complex Lorentz
transformations to the primitive domain of analyticity determined
by the spectral condition.
We recall~\cite{SW,BLOT} that $x\in \oR^{4n}$ belongs to the extended domain
if and only if the convex cone generated in $\oR^4$ by the points
 $\xi_k=x_k-x_{k+1}$, $k=1,\dots,n-1$, contains only spacelike vectors.
 In other words, the set of Jost points is the open cone
  $$
 {\cal J}_n=\left\{x\in \oR^{4n}:\quad \left(\sum_{k=1}^{n-1}
 \lambda_k(x_k-x_{k+1})\right)^2< 0\qquad \forall\, \lambda_k \geq 0,\,\,
 \sum_{k=1}^{n-1}\lambda_k > 0\right\}.
 \eqno{(12)}
 $$
 The covariance with respect to the action of the complex Lorentz group
 $L_+(\oC)$ implies the following transformation rule for the analytic
 Wightman functions under the total space-time inversion
  $$ {\cal W}_{\iota_1\dots
 \iota_n}(z_1,\dots,z_n)=(-1)^{2J}{\cal W}_{\iota_1\dots
 \iota_n}(-z_1,\dots,-z_n),
 \eqno{(13)}
  $$
 where $J$ is the total number of unpointed indices of the fields involved in
 the vacuum expectation value. If the expectation values grow in
   momentum space faster than exponentially of order 1, then
   the analyticity domain in coordinate space is empty because the Laplace
   transformation does not exist for such functions. However, as first
   observed by L\"ucke~\cite{L}, the symmetry (13) acts in a hidden manner
 even in this essentially nonlocal case and leads to important consequences.

{\bf 'heorem 3.} {\it Let $\{\phi_\iota\}$  be a family of fields
 defined on the test function space $S^0(\oR^4)$ and satisfying all Wightman
axioms except possibly locality. Assume that they transform according to the
irreducible representations $(j_\iota,k_\iota)$ of the group
$SL(2,\oC)$ and let  ${\cal W}_{\iota_1\dots \iota_n}$ be the generalized
function determined by the $n$-point vacuum expectation value
 $\langle\Psi_0,\, \phi_{\iota_1}(x_1)\dots
\phi_{\iota_n}(x_n)\Psi_0\rangle$. Then the complement $\complement{\cal J}_n$
 of Jost cone is a carrier of the functional
  $$
  {\cal  W}_{\iota_1\dots\iota_n}(x_1,\dots,x_n)- (-1)^{2J}{\cal
 W}_{\iota_1\dots\iota_n}(-x_1,\dots,-x_n),
  \eqno{(14)}
  $$
 where  $J=j_{\iota_1}+\dots+j_{\iota_n}$.

Proof.} We pass to the difference variables $\xi_k$ and to the generalized
 functions $W_{\iota_1\dots \iota_n}$ connected with
  ${\cal W}_{\iota_1\dots \iota_n}$ by the relation
  $$
 ({\cal W},\,f) =
 \left(W,\, \int\!f(t^{-1}\xi)\,{\rm d}\xi_n\right),
  \eqno{(15)}
 $$
 where $t:\,\,(x_1,\dots,x_n)\rightarrow
 (\xi_1=x_1-x_2,\dots,\xi_{n-1}=x_{n-1}-x_n,\xi_n=x_n)$.
 The correspondence   $W\to{\cal W}$
   defined by (15) is an injective mapping
 $S^{\prime 0}(\oR^{4(n-1)})\rightarrow S^{\prime 0}(\oR^{4n})$
 which is evidently continuous under the weak topologies of these
 spaces. It is easily verified that every translation-invariant functional
 belongs to its range. Furthermore, Lemma~4 in~\cite{S2} shows that
  $W \in S^{\prime 0}(U)$, with $U$  an open cone in
 $\oR^{4(n-1)}$, if and only if ${\cal W} \in S^{\prime 0}({\cal
 U})$, where ${\cal U}=\{x\in \oR^{4n}:  (x_1-x_2,\dots,x_{n-1}-x_n)\in U\}$.
  We regularize the ultraviolet behavior of
 $W$ by multiplying its  Fourier
transform $\check W$ with $\omega_M(p)=\omega((P\cdot P)/M^2)$, where
$P=\sum_{k=1}^n p_k$, the momentum-space variables $p_k$ are conjugates of
$\xi_k$, the inner product is Minkowskian, and
$\omega(t)$ is a smooth function with support in
the interval $(-1,1)$ and identically equal to 1 for $|t|\leq 1/2$.  Clearly $\omega_M$
is a multiplier for $S_0={\cal D}$ and, for every $u\in {\cal D}^\prime$,
 $u\,\omega_M$ tends  to $u$ in ${\cal D}^\prime$ as
$M\to \infty$.  From the spectral condition, it follows
   that $\check W_M=\check W\,\omega_M$ has a continuous extension to
   the Schwartz space $S$, i.e., is a tempered distribution. Namely,
   let us denote the closed forward light cone by $\overline{\oV}_+$  and
   prove the following lemma:

   {\bf Lemma 1}. {\it Let $u\in {\cal D}'(\oR^{4n})$  be a Lorentz-covariant
   distribution whose support is contained in the cone
    $\overline{\oV}_+\times\dots\times
    \overline{\oV}_+=\overline{\oV}_+^{\,n}$.  If $\omega\in {\cal D}(\oR)$
   and $P=\sum_{k=1}^n p_i$, then $u\,\omega(P\cdot P)\in
   S^\prime(\oR^{4n})$.

  Proof}. We denote by $T$ the representation according to which  $u$
  transforms under the action of $L^\uparrow_+$.  With a basis
  fixed in the representation space, $u$ can be identified with the system
  of distributions $u^i\in {\cal D}^\prime$; the number of these is equal to
  the dimension of the representation.
A distribution belonging to ${\cal D}'$ has a continuous extension to $S$ if
and only if its convolution with any test function $g\in{\cal D}$ has no
worse than a power growth at infinity. (This simple and convenient criterion
of extendability is proved in~\cite{Shi}, \S\,II.10.7.)
   The value of the convolution $(u^i\,\omega*g)$ at a
point $q$ is the value taken by the distribution $u^i$ on the shifted
function $g(p-q)$. We can assume, without loss of generality, that
  ${\rm supp}\,g$ lies in the set $\{p\in\oR^{4n}:  \,\|p\|<1/n\}$,
   where $\|\cdot\|$ is the Euclidean norm.  Let $Q=\sum_{k=1}^n q_k$.
  We need only consider the shifts along the
surface $Q\cdot Q=0$ because for the other directions,
   $(u^i\,\omega*g)(q)$  vanishes for
sufficiently large $\|q\|$. It can be additionally assumed that $Q^2=Q^3=0$,
because any vector in $\oR^4$ is converted to this form
by an appropriate spatial rotation.
 We now use the light-cone variables $Q^\pm=(Q^0\pm  Q^1)/\sqrt{2}$
and set $Q^-=0$, $Q^+\to+\infty$ for definiteness. Let $\Lambda$ be
the  transformation $p^+_k\to p^+_k/Q^+$, $p^-_k\to  Q^+p^-_k$, $k=1,\dots,n$.
 In view of the Lorentz covariance of $u$ and invariance of $\omega$, we have
     $$
    (u^i\,\omega *g)(q)=\sum_j
   T^i_j(\Lambda)(u^j,g_q),\quad {\rm where} \quad g_q(p)= \omega(P\cdot
   P)\,g(q-\Lambda^{-1}p).
   \eqno{(16)}
    $$
   The points of ${\rm supp}\, g_q$ satisfy the inequalities
   $|P\cdot P|<1$ and $(P^2)^2+(P^3)^2<1$
 by construction and hence  $|P^+P^-|<1$. Furthermore $|Q^+-Q^+P^+|<1$.
 Therefore,  if $Q^+$ is sufficiently large, ${\rm  supp}\,g_q$ is contained
 in the set $\|P\|<2$.
 We fix a neighborhood  ${\cal V}$
     of the support of $u$ by taking the union of a
    neighborhood of the origin with the product $V^n$, where $V$ is an open
    properly convex cone in $\oR^4$ containing
  $\overline \oV_+\setminus \{0\}$.
      For the points of $V^n$, the inequality $\|p\|<\theta \|P\|$
    holds with some constant $\theta>0$, because otherwise we could find a
    sequence of points $p_{(\nu)}\in V^n$ such that $\|p_{(\nu)}\|=1$ and
    $\|P_{(\nu)}\|<1/\nu$.  Then we could choose a convergent subsequence
    whose limit $\bar p$ is a nonzero vector in  $\overline{V}^{\,n}$ such
    that $|\bar P|=0$, which contradicts the assumption that the cone $V$ is
    properly convex.  Therefore, the set ${\rm supp}\,g_q\cap  {\cal V}$ lies
    in the ball of radius $2\theta$ and we have the estimate
     $$
    |(u^j,\,g_q)|\leq C_j\|g_q\|_{2\theta,N_j},
    \eqno{(17)}
    $$ where
 $\|g_q\|_{2\theta,N}=\max_{|\kappa|\leq N}\, \sup_{\|p\|\leq
    2\theta}|\partial^{\,\kappa} g_q(p)|$ and $N_j$ has the meaning of the
singularity order of the distribution $u^j$ in this ball. The
   transformation $\Lambda^{-1}$ contracts the graph of $g$ by $Q^+$ times
     with respect to every variable $p^+_k$. Therefore,
     $$
     \sup_p
    |\partial^{\,\kappa} g(q-\Lambda^{-1}p)|=\sup_p|\partial^{\,\kappa}
   g(\Lambda^{-1}p)|\leq C_\kappa {Q^+}^{|\kappa|}
    $$
     and, consequently,
     $$
     \|g_q\|_{2\theta,N}\leq C_{N}\,(1+\|q\|)^{N}.
      \eqno{(18)}
    $$
     Taking into account that the representation
     matrix elements $T^j_k(\Lambda)$ are rational functions of
      the boost parameter $Q^+$ and  combining  (16)--(18),
      we conclude that the function  $(u^i\omega*g)(q)$ is
       polynomially bounded. Lemma~1 is thus proved.

   We return to the proof of Theorem 3 and consider the (inverse)
 Laplace transform  ${\bf W}_M$ of the distribution  ${\check
 W}_M$.  It is holomorphic on the usual tube
  $\oT_{n-1}=\oR^{4(n-1)}-i\oV_+^{(n-1)}$ and $W_M$ is its boundary value.
Since the regularization preserves the Lorentz covariance, we can apply
  the Bargman-Hall-Wightman theorem~\cite{SW, BLOT},
   which shows that ${\bf W}_M$  allows an analytic continuation into
 the extended domain $\oT_{n-1}^{\,\rm ext}$ and the continued function is
 covariant under the complex Lorentz group $L_+(\oC)$.
Therefore, the function ${\cal
 W}_M(z_1,\dots,z_n)={\bf W}_M(z_1-z_2,\dots, z_{n-1}-z_n)$ satisfies
   (13) in the corresponding analyticity domain. As a consequence, the
  tempered distribution
  $$
  F_M\stackrel{\rm def}{=} {\cal
W}_M(x_1,\dots,x_n)-(-1)^{2J} {\cal W}_M(-x_1,\dots,x_n)
  $$
  vanishes in the
  Jost cone and its restriction to $S^0$ is carried by $\complement{\cal
J}_n$. Moreover, it has a continuous extension to the space $S^0(U)$
associated with the open cone $U=\complement\overline {\cal J}_n$, i.e., with
  the interior of the cone complementary to the Jost cone.  In fact, this
   extension $\tilde F_M$ can be defined by  $(\tilde
   F_M,f)=(F_M,\chi f)$, where $\chi$ is a multiplier for the Schwartz space
 $S$  which is identically equal to 1 in an $\epsilon$-neighborhood of
   $\complement{\cal J}_n$ and vanishes outside the $2\epsilon$-neighborhood.
   Such a multiplier satisfies the estimate $|\partial^q\chi(x)|\le C
  h^{|q|}$ and, for any function $f \in S^0(U)$, we have the inequalities
 $|\partial^qf(x)|\le C'\,\|f\|_{b,N}\,b^{|q|}(1+\|x\|)^{-N}$ which hold on
  ${\rm supp}\,\chi$.
  Hence, the multiplication by $\chi$ continuously maps $S^0(U)$ into $S$.
   It is  important that the extensions  $\tilde F_M$ are compatible with each
 other if $M$ and $M'$ are large enough compared to $b$, namely,
   $$
  {\tilde
 F}_M|S^{0,\,b}(U)= {\tilde F}_{M'}|S^{0,\,b}(U).
  \eqno{(19)}
  $$
   To prove (19), we use the density theorem mentioned in Section 2.
    Its more detailed formulation given in~\cite{S1,S4}
     shows that there is a constant $c$
    such that for $b^\prime\geq cb$, the space $S^{0,\,b^\prime}$ is dense in
 $S^{0,\,b}(U)$ in the topology of $S^{0,\,b^\prime}(U)$. Let $M,
   M^\prime>2cb$, $f\in S^{0,\,b}(U)$, $f_\nu\in S^{0,\,cb}$, and let $f_\nu\to
   f$ in $S^{0,\,cb}(U)$.  Then $f_\nu\in S^{0,\,M/2}\cap S^{0,\,M'/2}$ and
   we have the estimate $|f_\nu(z)|\le C_{\nu,
   N}e^{\min(M,M')\|y\|/2}(1+\|x\|)^{-N}$, which implies that
   ${\rm supp}\,\check f_\nu$ are contained in the ball $\|p\|\le
   \min(M,M')/2$, where both regularizing multipliers $\omega_M$,
   $\omega_{M'}$ are equal to $1$ by construction. Therefore, $({\tilde
   F}_M,f)= ({\tilde F}_{M'},f)=\lim_{\nu\to\infty}(F,f_\nu)$, where $F=
   {\cal W}(x)-(-1)^{2J}{\cal W}(-x)$. Thus, there exists a continuous
   extension of the functional $F$ to $S^0(U)$. This completes the proof.

\section{Generalization of the PCT theorem}

  We now  formulate an analog of the Jost-Dyson weak locality
  condition.

   {\bf Definition 2}.   Let $\{\phi_\iota\}$ be a family of quantum fields
   defined on the test function space $S^0(\oR^4)$, and with a common
   invariant domain in the Hilbert space of states.  We say that
   $\{\phi_\iota\}$ satisfies the weak relative asymptotic commutativity
   condition if for each system of indices  $\iota_1,\dots,\iota_n$, the
   functional $$ \langle\Psi_0,\,
   \phi_{\iota_1}(x_1)\dots\phi_{\iota_n}(x_n)\Psi_0\rangle
      -i^F\langle\Psi_0,\,
  \phi_{\iota_n}(x_n)\dots\phi_{\iota_1}(x_1)\Psi_0\rangle,
   \eqno{(20)}
   $$
  where $F$ is the number of fields with half-integer spin in the monomial
  $\phi_{\iota_1}\dots\phi_{\iota_n}$, is carried by the complement
   $\complement{\cal J}_n$ of Jost cone.

This condition is certainly fulfilled in theories with the normal asymptotic
commutation relations because Jost points are totally spacelike, i.e,
 $(x_1,\ldots,x_n)\in {\cal J}_n$ implies $(x_j-x_k)^2<0$  for all $j\ne k$.

{\bf Theorem 4}. {\it  Assume we are dealing with quantum fields
$\{\phi_\iota\}$ defined on $S^0(\oR^4)$ and satisfying all Wightman axioms
 except possibly locality. Let $\phi_\iota$ transform according to the
irreducible representation $(j_\iota,k_\iota)$ of the group $SL(2,\oC)$.
Then the weak relative asymptotic commutativity condition is equivalent to
 the existence of an antiunitary $PCT$-symmetry operator $\Theta$ which
 leaves the vacuum invariant and acts on the fields according to the rule
  $$
 \Theta\,\phi_\iota(x)\,\Theta^{-1}=
       (-1)^{2j_\iota}\,i^{F_\iota}\,\phi_\iota(-x)^*,
       \eqno{(21)}
       $$
       where $F_\iota$ is the spin number of $\phi_\iota$.

Proof.} Let the $PCT$ symmetry hold and an operator $\Theta$ with the listed
properties exist. Let us consider the vacuum expectation value of the
monomial $\phi_{\iota_1}(x_1)\cdots\phi_{\iota_n}(x_n)$. Applying
(21) and using the invariance of $\Psi_0$ and the antiunitary property of
$\Theta^{-1}$, we obtain the relation
       $$
       {\cal W}_{\iota_1\dots
      \iota_n}(x_1,\dots,x_n)=(-1)^{2J}\,i^F\,{\cal W}_{\iota_n\dots
     \iota_1}(-x_n,\dots,-x_1), \eqno{(22)}
      $$
      where
      $J=j_{\iota_1}+\dots+j_{\iota_n}$, $F=F_{\iota_1}+\dots+F_{\iota_n}$.
      In deriving (22), it should be kept in mind that the expectation value
      is zero for odd $F$ and that  $(-i)^F=i^F$ for even $F$. Subtracting the
      functional $(-1)^{2J}\,{\cal W}_{\iota_1\dots\iota_n}(-x_1,\dots,-x_n)$
   from the left-hand and right-hand sides of (22) and applying Theorem 3,
   we conclude that the  weak relative asymptotic commutativity condition is
   fulfilled. Conversely, if $\{\phi_\iota\}$ satisfies this condition, then
      the difference
       ${\cal W}_{\iota_1\dots\iota_n}(x_1,\dots,x_n)-
      (-1)^{2J}\,i^F\,{\cal W}_{\iota_n\dots\iota_1}(-x_n,\dots,-x_1)$
     is representable as a sum of two functionals carried by the cone
     $\complement{\cal J}_n\ne \oR^{4n}$. Its Fourier transform has support
      in the properly convex cone
       $$
      \left\{p\in \oR^{4n}:\,\sum_{k=1}^n
      p_k=0,\quad\sum_{k=1}^lp_k\in \overline{\oV}_+,\quad
      l=1,\dots,n-1\right\}
      $$
       determined by the spectral condition.
      Therefore, equality (22) holds identically by Theorem 1. The operator
      $\Theta$ can now be constructed in the ordinary way.
   First we define it on those vectors that are obtained by applying
    monomials in fields to the vacuum. Namely, we set
       $$
   \Theta\Psi_0=\Psi_0,\quad
  \Theta\phi_{\iota_1}(f_1)\dots\phi_{\iota_n}(f_n)\Psi_0
  =(-1)^{2J}\,i^F\,
  \phi_{\iota_1}(f^-_1)^*\dots\phi_{\iota_n}(f^-_n)^*\Psi_0,
  $$
  where $f^-(x)=f(-x)$.
    It easily seen that $\Theta$ is well defined.
  In fact, taking into account
    that  $\phi^*_\iota$ transforms according to the
    conjugate representation $(k_\iota,j_\iota)$, we see that (22) implies
  the relation
  $\langle\Theta\Phi,\,\Theta\Psi\rangle=\overline{\langle\Phi,\,\Psi\rangle}$
   for vectors of this special form. Therefore, if a vector $\Psi$
  is generated by different monomials $M_1(f_1),\,M_2(f_2)$, then the scalar
  product $\langle\Theta M_1\Psi_0,\,\Theta  M_2\Psi_0\rangle$ is equal
  to squared length of each of the vectors
  $\Theta M_1\Psi_0,\,\Theta M_2\Psi_0$, i.e., these vectors coincide.
  Analogously, if $\Psi=\Psi_1+\Psi_2$, where all vectors are obtained by
  applying monomials to $\Psi_0$, then  $\Theta\Psi=\Theta\Psi_1+\Theta\Psi_2$.
  Therefore, $\Theta$ can be extended to $D_0$ by antilinearity.   A further
  extension by continuity yields an antiunitary operator on ${\cal H}$.
  This completes the proof.

  Theorem 15 of~\cite{S2} shows that the asymptotic commutativity condition
  stated in Section~2 guarantees the existence of a Klein transformation
  reducing the commutation relations to the normal form. Because of this, the
  $PCT$  symmetry holds in the nonlocal theories satisfying this condition.
  The proof given above shows also that this symmetry  holds even if the
  difference (20) is carried by the complement of a cone generated by an
  arbitrarily small real neighborhood of a Jost point and that then this
  functional is necessarily  carried by $\complement{\cal J}_n$.

\section{Transitivity of weak relative asymptotic commutativity}

{\bf Theorem 5}. {\it Let $\{\phi_\iota\}$ be a family of fields
satisfying the assumptions of Theorem 4 and $\Theta$ be the
corresponding $PCT$-symmetry operator. Let $\psi$ be
 another field with the same domain of definition  and transforming according
 to the representation $(j,k)$ of  $SL(2,\oC)$. Suppose that the joined
 family $\{\phi_\iota,\, \psi\}$ satisfies all Wightman axioms except
 locality and that for any system $\iota_1,\dots,\iota_n$ of indices and for
 every $m$, the functionals $$ \langle\Psi_0,\,
\phi_{\iota_1}(x_1)\dots\phi_{\iota_m}(x_m)\psi(x)
\phi_{\iota_{m+1}}(x_{m+1})\dots\phi_{\iota_n}(x_n)\Psi_0\rangle
$$
$$
  {}-i^F\langle\Psi_0,\,
  \phi_{\iota_n}(x_n)\dots\phi_{\iota_{m+1}}(x_{m+1})
\psi(x)\phi_{\iota_m}(x_m)\dots\phi_{\iota_1}(x_1)\Psi_0\rangle,
   \eqno{(23)}
   $$
  where $F$ is the number of spinor fields in the monomial
  $\phi_{\iota_1}\cdots\phi_{\iota_n}\psi$, are carried by the cone
   $\complement{\cal J}_{n+1}$. Then  $\Theta$ implements
   the $PCT$ symmetry for $\psi$ as well and the joined family
  of fields satisfies the weak asymptotic commutativity condition.

  Proof.} Applying Theorems 1 and 3 as above, we find that
 $$
  \langle\Psi_0,\, \phi_{\iota_1}(x_1)\dots\phi_{\iota_m}(x_m)\psi(x)
\phi_{\iota_{m+1}}(x_{m+1})\dots\phi_{\iota_n}(x_n)\Psi_0\rangle=
$$
$$
  {}=(-1)^{2J}\,i^F\langle\Psi_0,\,
  \phi_{\iota_n}(-x_n)\dots\phi_{\iota_{m+1}}(-x_{m+1})
\psi(-x)\phi_{\iota_m}(-x_m)\dots\phi_{\iota_1}(-x_1)\Psi_0\rangle,
   \eqno{(24)}
   $$
  where $J=j_{\iota_1}+\dots+j_{\iota_n}+j$. After averaging with the test
  function $f(x_1)\dots f(x_n)f(x)$, the relation (24) takes the form
   $$
  \langle\Theta^{-1}\Phi,\,\psi(f)\Psi\rangle=
  (-1)^{2j}\,(-i)^{F(\psi)}\langle\Theta\Psi,\,\psi(f^-)\Phi\rangle,
  \eqno{(25)}
  $$
  where the following designations are used:
   $$
   \Phi=
  \phi_{\iota_m}(f^-_m)\dots\phi_{\iota_1}(f^-_1)\Psi_0,\quad
  \Psi=\phi_{\iota_{m+1}}(f_{m+1})\dots\phi_{\iota_n}(f_n)\Psi_0.
   \eqno{(26)}
  $$
  Performing the complex conjugation and using the antiunitary property of
   $\Theta$, we obtain
   $$
  \langle\Phi,\,\Theta\psi(f)\Psi\rangle=
  (-1)^{2j}\,i^{F(\psi)}\langle\Phi,\,\psi(f^-)^*\Theta\Psi\rangle,
  $$
  Because of the cyclicity of the vacuum with respect to $\{\phi_\iota\}$,
 the subspaces spanned by vectors $\Phi$ and $\Psi$ of the form (26) are dense
  in the Hilbert space, and we conclude that
   $$
   \Theta\,\psi(f)\,\Theta^{-1}=
  (-1)^{2j}\,i^{F_\psi}\,\phi(f^-)^*.
   \eqno{(27)}
   $$
Thus, the operator $\Theta$ transforms $\psi$ correctly, as was to be proved.

{\bf Corollary}.  The weak relative asymptotic commutativity property is
transitive in the sense that if each of fields $\psi_1, \psi_2$
satisfies the assumptions of Theorem 5, then this property holds for
 $\{\psi_1,\psi_2\}$.

Indeed, then there is a $PCT$-symmetry operator common to the fields
$\{\phi_\iota,\, \psi_1,\,\psi_2\}$ and by Theorem 4,
the weak relative asymptotic commutativity  condition is satisfied not only
for $\{\psi_1,\psi_2\}$ but also for the whole family
$\{\phi_\iota,\psi_1,\psi_2\}$.

The developed technique can also be used in deriving the transitivity of
relative locality in  hyperfunction QFT. Specifically, let us prove the
following proposition.

 Let $\{\phi_\iota\}$ be a family of local quantum fields defined on the test
 function space $S^1_1(\oR^4)$, $D_0$ be its cyclic domain in the
 Hilbert space, and assume the standard spin-statistics relation holds.
  Let $\{\psi_1,\psi_2\}$ be a pair of fields with the same domain of
 definition and such that the joined family $\{\phi_\iota,\psi_1,\psi_2\}$
 satisfies all Wightman axioms except possibly locality which is replaced by
 the assumption that
  $$
[\phi_\iota(x),\psi_1(x')]_{\stackrel{-}{(+)}}\Psi=0,\quad
[\phi_\iota(x),\psi_2(x')]_{\stackrel{-}{(+)}}\Psi=0
\quad {\rm for}\,\,
(x-x')^2<0, \,\, \Psi\in D_0,
\eqno{(28)}
$$
where the commutation relations are normal. Then
$$
[\psi_1(x),\psi_2(x')]_{\stackrel{-}{(+)}}\Psi=0
\quad {\rm for}\,\,
(x-x')^2<0, \,\, \Psi\in D_0,
\eqno{(29)}
$$
and this commutation relation is also normal.

In fact, let us consider the vacuum expectation value
 $$ \langle\Psi_0,\,
\phi_{\iota_1}(x_1)\dots\phi_{\iota_m}(x_m)\psi_1(x)\psi_2(x')
\phi_{\iota_{m+1}}(x_{m+1})\dots\phi_{\iota_n}(x_n)\Psi_0\rangle
\eqno{(30)}
$$
at real points of analyticity
$(x_1,\ldots,x_m,x,x',x_{m+1}\ldots, x_n)\in {\cal J}_{n+2}$.
Proposition 9.15 in~\cite{BLOT} showing the transitivity of weak relative
locality for tempered distribution fields is directly extendable to Fourier
hyperfunction QFT. Therefore,
the fields $\phi_\iota,\psi_1,\psi_2$ are weakly relatively local  and
 we can invert the order of field operators in (30), which results in
 appearance of the factor $i^F$, where $F$ is the number of spinor fields
 in the monomial under consideration.
 Next we use locality of $\phi_\iota$ and the assumption of relative locality
 (28) to restore the initial order of field operators in the expectation
 value. Then the factor is changed to $-1$ if both fields $\psi_1,\psi_2$
 have half-integer spin and to $1$ in the other cases. So we have
  $$ \langle\Psi_0,\, \phi_{\iota_1}(x_1)\dots\phi_{\iota_m}(x_m)
 [\psi_1(x),\,\psi_2(x')]_{\stackrel{-}{(+)}}
\phi_{\iota_{m+1}}(x_{m+1})\dots\phi_{\iota_n}(x_n)\Psi_0\rangle=0
\eqno{(31)}
$$
 for all Jost points.
 In momentum space, the generalized function (31) has support in the wedge
  determined by
 $$
      \sum_{k=1}^lp_k\in \overline{\oV}_+,\quad l=1,\dots,m;\quad
      \sum_{k=1}^np_{m+k}\in \overline{\oV}_-,\quad l=1,\dots,n-m.
  $$
 It remains to note that any pair of points $(x,x')$ such that
 $(x-x')^2<0$ enters in a Jost point
 $(x_1,\ldots,x_m,x,x',x_{m+1}\ldots, x_n)\in {\cal J}_{n+2}$ and apply
Theorem 2.

 \section{S--equivalence of nonlocal fields}

 The Haag-Ruelle scattering theory shows that the existence of S--matrix is
derivable from  the general principles of local QFT complemented by some
technical assumptions about the  energy momentum spectrum  and the
structure of the single particle space, see~\cite{BLOT}.
Under these assumptions, we can assign to each field
$\phi_\iota$ an operator $\varphi_\iota(g,t)$ depending on the parameter
 $t\in \oR$ and the function $g({\bf p})\in {\cal D}(\oR^3)$ so that in the
Hilbert space of states, there exist the strong limits
 $$
\Phi^{\stackrel{{\rm
in}}{{\rm out}}}_{\iota_1,\dots,\iota_n}(g_1,\dots,g_n)=
\lim_{t\to\mp\infty}\varphi_{\iota_1}(g_1,t)\cdots
\varphi_{\iota_n}(g_n,t)\Psi_0,
\eqno{(32)}
$$
describing  the incoming and outgoing scattering states of $n$ particles of
the kind $\iota_1,\dots,\iota_n$  with the momentum space wave packets
  $g_1,\dots,g_n$. Moreover, the linear operators $\phi_\iota^{{\rm ex}}(g)$
(${\rm ex}={\rm in},{\rm out}$) determined by
  $$
  \phi_\iota^{{\rm
ex}}(g) \Phi^{{\rm ex}}_{\iota_1,\dots,\iota_n}(g_1,\dots,g_n)= \Phi^{{\rm
 ex}}_{\iota,\iota_1,\dots,\iota_n}(g,g_1,\dots,g_n),
  \eqno{(33)}
 $$
 with the subsequent extension by linearity, are well defined and represent
free fields averaged with the corresponding test functions.
In this construction, the local commutativity axiom is used only in deriving
the strong cluster decomposition property of vacuum expectation values.
 This property  plays the central role in the proof of the existence of
limits (32).  However, the asymptotic commutativity condition ensures
 the strong cluster property as well if the usual assumption is made that
the theory has a finite mass gap between the one-particle state and the
continuum.  Namely, if ${\cal W}^T_n$ is the truncated part of the
$n$--point vacuum expectation value and $f\in S^0(\oR^{4n})$, then the
convolution $({\cal W}^T_n*f)(x)$ considered at equal times as a function of
the difference variables ${\bf x}_i-{\bf x}_j$ belongs to the space
$S^0(\oR^{3(n-1)})$.  Because of this, the Haag--Ruelle construction
admits a direct generalization to the nonlocal theories under study,
 see~\cite{BL,FS}.  The scattering matrix is defined by
  $$ S\Phi^{{\rm
out}}_{\iota_1,\dots,\iota_n}(g_1,\dots,g_n)= \Phi^{{\rm
in}}_{\iota_1,\dots,\iota_n}(g_1,\dots,g_n)
  $$
 and is an isometric operator with the domain ${\cal H}^{{\rm  out}}$ and
  the range  ${\cal H}^{{\rm  in}}$, where ${\cal H}^{{\rm  ex}}$ are  the
  closures of linear spans $D^{{\rm ex}}$ of vectors
  (32) and $\Psi_0$.  The asymptotic free fields are connected by the relation
   $$
  \phi_\iota^{{\rm out}}(x)=S^{-1}\phi_\iota^{{\rm
  in}}(x)S.
   \eqno{(33)}
  $$
   The results of Section 6 show that Theorem 4.20 in~\cite{SW} on  the
 $S$--equivalence of quantum fields can be generalized in the following way:

  {\bf Theorem 6}. {\it Let $\phi_1(x)$ be a field
defined on the space $S^0(\oR^4)$ and satisfying all
Wightman axioms  with  the asymptotic commutativity substituted
for locality.  Assume that a field $\phi_2(x)$ transforms according to the
  same representation of the Lorentz group and the pair $\{\phi_1, \phi_2\}$
  satisfies the weak asymptotic commutativity condition.
  If the  ``in'' and ``out'' limits exist and
   $\phi_1^{{\rm in}}(x)=\phi_2^{{\rm in}}(x)$, then
  $\phi_1^{{\rm out}}(x)=\phi_2^{{\rm out}}(x)$.

  Proof.}  Indeed, by Theorem 5 the fields $\phi$ and $\psi$
  have a $PCT$ operator which transforms
  ``in'' fields into ``out'' fields according to the formula
  $$
        \Theta\,\phi_\iota^{{\rm in}}(x)\,\Theta^{-1}=
       (-1)^{2j_\iota}\,i^{F_\iota}\,\phi_\iota^{{\rm out}}(-x)^*,
       \eqno{(34)}
  $$
  precisely as in local field theory.

  In particular, a field $\phi(x)$ with nontrivial S--matrix
        cannot be weakly asymptotically commuting with its
  associated free fields  $\phi^{{\rm in}}(x)$ and $\phi^{{\rm out}}(x)$.

  If in addition the asymptotic completeness ${\cal H}^{{\rm  ex}}=
  {\cal H}$ is assumed, then $S$--matrix can be expressed through the
operator  $\Theta$ and the $PCT$ operator of  the free fields. Specifically,
  applying  $\Theta_{{\rm out}}$ to  (34) and using the relation
  $$
         \Theta_{{\rm out}}^{-1}\,\phi_\iota^{{\rm out}}(-x)^*\,\Theta_{{\rm
        out}}= (-1)^{2j_\iota}\,(-i)^{F_\iota}\,\phi_\iota^{{\rm
  out}}(x),
        $$
        we obtain
       $$
        S= \Theta^{-1}\Theta_{{\rm out}}.
       \eqno{(35)}
  $$
  The same result follows directly from the definition~\cite{BLOT}
  of action of the operators $\Theta$ and $\Theta_{{\rm ex}}$
  on  $D^{{\rm ex}}$.  Relation (35) implies the unitary property of
  the $S$--matrix and, taking into account the equality $\Theta^2=\Theta_{{\rm
  out}}^2$, the $PCT$ invariance:
    $$
   \Theta_{{\rm out}}S\,\Theta_{{\rm
  out}}^{-1}=S^*.
   $$
  Thus, the asymptotic commutativity condition stated in
  Section 2 guarantees  the fulfillment of all properties required for the
  physical interpretation of nonlocal QFT and established
  previously by the Haag--Ruelle theory for local fields.

  \section{Concluding remarks}

    The Borchers class of a  massive scalar free field $\phi$
  in three and more dimensional space-time was first
  determined by Epstein~\cite{E} under the assumption that the
  vacuum expectation values are tempered distributions, and it turned out
  to be just the Wick polynomials including derivatives of
  $\phi$.  Under the same assumption of temperedness, Baumann~\cite{B} has
  proved that if a massless scalar field in $3+1$ dimensional space-time has
  a trivial S--matrix, then this field is relatively local to the free field.
  In~\cite{NM}, infinite series of Wick powers of a massive free field were
  studied. These series were shown to be well defined
  in the sense of hyperfunction and to extend the Borchers class in the same
  sense if they are of order $<2$ or of order 2 and type 0. The Wick--ordered
  entire functions  with an arbitrary order of growth are
  considered in~\cite{SS}. A relatively straightforward application of the
  Cauchy-Poincar\'e theorem shows that these functions satisfy the asymptotic
  commutativity condition in both the massive and massless cases and
  for any space-time dimension.
    It is natural to suppose that the convergent series of Wick powers acted
  upon by differential operators of an infinite order exhaust the
  nonlocal extension of Borchers class of free field.

  In order to construct the scattering states and asymptotic free fields, it
  suffices to use the above-mentioned version of cluster property which
  means that the truncated Wightman functions decrease faster than any
  inverse power of the variable $\max_{i,j}|{\bf x}_i-{\bf x}_j|$.
  As shown in~\cite{S7}, this decrease is actually exponential in theories
  with a mass gap regardless their local properties, and this is essential
  to the derivation of bounds on growth of  scattering amplitudes in
  nonlocal QFT's.  A serious difficulty in dealing with analytic functionals
  is that the limit of functionals carried by a closed cone is not
  necessarily carried by the same cone, in contrast to the customary notion
  of support of a distribution. Because of this, the domains of nonlocal
  field operators in the Hilbert space need much more attention compared to
  the standard local QFT.  It should be emphasized that questions of this
  kind are important even for formulating hyperfunction QFT, where the
  closure of Hermitian field operators can destroy localization properties,
  see~\cite{BN2}. In this paper, we have made no attempt to derive an analog
  of Theorem 2 for $u\in{\cal D}^\prime$, which is desirable to accomplish
   the suggested generalization of Borchers classes.  It would be also
  worthwhile to consider possible links of the developed scheme with the
  modern derivation of the spin-statistics  and $PCT$ theorems in the
  framework of algebraic QFT~\cite{BY,Mu}, where the modular covariance plays
  a significant role and the localization of charges in spacelike cones
  instead of in compact regions is admitted.

{\bf Acknowledgments.}

  The author would like to thank Professor V.~Ya.~Fainberg for interesting
  discussions.
  He is also grateful to the Russian Foundation for Basic Research for
   financial support under Contract 99-02-17916 and Grant 00-15-96566.

 \end{document}